\documentclass[pr,twocolumn,showpacs,superscriptaddress,10pt]{revtex4-1}
\usepackage[colorlinks,citecolor=blue,linkcolor=red,dvipdfm]{hyperref}
\usepackage{amsmath,amssymb,graphicx}

\begin{document}

\title{Collisions and fusion of one- and two-dimensional solitons driven by potential troughs in the cubic-quintic nonlinear Schr\"{o}dinger equations}

\author{Liangwei Zeng}
\affiliation{School of Arts and Sciences, Guangzhou Maritime University, Guangzhou 510725, China}
\affiliation{College of Physics and Optoelectronic Engineering, Shenzhen University, Shenzhen 518060, China}

\author{Boris A. Malomed}
\affiliation{Department of Physical Electronics, School of Electrical Engineering, Faculty of Engineering,
and Center for Light-Matter Interaction, Tel Aviv University, P.O.B. 39040, Tel Aviv, Israel}
\affiliation{Instituto de Alta Investigaci\'{o}n, Universidad de Tarapac\'{a}, Casilla 7D, Arica, Chile}

\author{Dumitru Mihalache}
\affiliation{Horia Hulubei National Institute of Physics and Nuclear Engineering, 077125 Magurele, Bucharest, Romania}

\author{Jingzhen Li}
\affiliation{College of Physics and Optoelectronic Engineering, Shenzhen University, Shenzhen 518060, China}

\author{Xing Zhu}
\email{\underline{xingzhu@gzmtu.edu.cn}}
\affiliation{School of Arts and Sciences, Guangzhou Maritime University, Guangzhou 510725, China}

\begin{abstract}
We study the formation and collision of one- and two-dimensional (1D and 2D) Gaussian-shaped and flat-top (FT) solitons in the framework of the nonlinear Schr\"{o}dinger equation with the cubic-quintic nonlinearity and two intersecting potential troughs. We find that Gaussian-Gaussian and Gaussian-FT collisions between the solitons, steered by the troughs, are quasi-elastic, while the collisions between FT solitons may be either quasi-elastic or inelastic, in the form of merger into a single FT soliton, thus spontaneously breaking the symmetry between the steering troughs. The Gaussian-FT collisions, being overall quasi-elastic, generate weak radiation fields.
\end{abstract}

\maketitle

The interactions of two or more components for solitons are interesting and have attracted many attentions in recent few decades. Specifically, the collisions of optical solitons have gained lots of interest due to their potential application in the all-optical information processing. Generally speaking, there are two types of collisions on solitons, namely, they are elastic or inelastic. Figuring out the physical schemes of such two types of collisions will be helpful for the better understanding of interactions of solitons. Despite the collisions of solitons in various physical setting have been reported, the collisions of flat-top solitons have not been investigated. The flat-top soliton is an interesting type of soliton families whose shape is quite different from the common solitons. In other words, the flat-top solitons can be written as a funtion of super gauss, while the common solitons are always described by the function of gauss. Therefore, the collisions of flat-top solitons need more attentions.

\section{Introduction}

The fundamental significance of solitons, supported by the balance of
dispersion/diffraction and nonlinearity, is commonly known \cite%
{Zakh,Ablowitz,Newell}, including their applications to various branches of
physics \cite{REV1,Dauxois,REV2,REV3,REV4}. While a majority of works in this area
addressed one-dimensional (1D) solitons, the study of 2D and 3D ones is a
challenging topic, because of their propensities to develop instability \cite%
{CLP1,CLP2,CLP3,REV5,R1,R2,R3}. In particular, the creation and stabilization of
multidimensional solitons is facilitated by the application of linear \cite%
{Peli,LP1,LP1b,LP2,LP3} and nonlinear \cite{NP1,NP2,NP3} potentials, as well as by
the resort to the cubic-quintic (CQ) \cite{CQ1,CQ2,CQ3,CQ3b,CQ3c,CQ4} and saturable
nonlinearities \cite{SATU1,SATU2,SATU3,SATU4,SATU5}. The use of the CQ and higher-order
nonlinear response of the material medium is a promising technique in
optics, where the desirable nonlinearity can be accurately engineered in
composite materials, such as colloidal suspensions of metallic nanoparticles
\cite{Cid1,Cid2}.

Interactions between solitons are an important aspect of the theoretical and
experimental work with solitons in various settings \cite{INT1,INT2,INT3,INT4}.
Such processes may be helpful to the transformation of the states of
solitons. Specifically, collisions \cite{CLS1} of solitons have been
reported in many works \cite{CLS1,CLS6,CLS2,CLS7,CLS3,CLS4,CLS5}. Most previous works focused
on collisions of tightly localized (Gaussian-shaped) solitons. On the other
hands, systems including competing nonlinearities, such as CQ combinations,
produce solitons families combining Gaussian-like modes and extended
flat-top (FT) ones \cite{Bulgaria,Michinel1,Michinel2,FTS0,FTS1,FTS2,FTS3}. In this work, we focus on
the systematic numerical analysis of collisions of traveling solitons in 1D
and 2D models with the CQ nonlinearity, including the collisions between
Gaussians, FT solitons, as well as Gaussian-FT collisions. The propagation
of the solitons is steered by potential troughs, which is a natural setting
in optics, where the troughs can be readily built by patterning the local
structure of the refractive index. The outputs of the collisions, such as
elastic passage or fusion of solitons into a single mode, can be used for
the design of optical devices, such as all-optical data-processing schemes.

\section{The model}

The propagation of light beams in optical media with the CQ nonlinearity is
modeled by the well-known nonlinear Schr\"{o}dinger equation, written in the
scaled form:%
\begin{equation}
i\frac{\partial E}{\partial z}=-\frac{1}{2}\nabla ^{2}E+V\left( x,y\right)
E-\left\vert E\right\vert ^{2}E+\left\vert E\right\vert ^{4}E.  \label{NLSE}
\end{equation}%
Here, $E$ and $z$ denote the field amplitude and propagation distance,
respectively, $\nabla ^{2}=\partial ^{2}/\partial x^{2}$ or $\nabla
^{2}=\partial ^{2}/\partial x^{2}+\partial ^{2}/\partial y^{2}$ are the
operators of the paraxial diffraction in the 1D and 2D settings,
respectively, while the cubic and quintic terms account for the
self-focusing and defocusing nonlinearity, respectively.

Equation (\ref{NLSE}) includes the steering potential troughs whose 1D form
can be written as
\begin{equation}
V_{\mathrm{1D}}(x)=-A\sum_{i=1,2}\mathrm{sech}[V_{i}(x-x_{i})^{2}],
\label{VE1D}
\end{equation}%
where $A>0$ and $V_{i}$ ate the troughs' depth and inverse width, and
\begin{equation}
x_{1}=\tilde{x}_{1}-\Omega _{1}z,x_{2}=\tilde{x}_{2}+\Omega _{2}z,
\label{Omega}
\end{equation}%
with constants $\tilde{x}_{1,2}$ and $\Omega _{1,2}$, the latter
coefficients defining the tilt of the troughs in the $\left( x,z\right) $
plane. In the 2D setting, the steering potential $V$ is composed of
cylindrical conduits,%
\begin{equation}
V_{\mathrm{2D}}=-A\sum_{i=1,2}\mathrm{sech}\{V_{i}[(x-x_{i})^{2}+y^{2}]\},
\label{VE2D}
\end{equation}%
which are determined by the circles in the $\left( x,y\right) $ plane and
tilted in the $\left( x,z\right) $ plane.

In this work, we first look for the stationary solutions of Eq. (\ref{NLSE}%
), with real propagation constant $b$, as $E(x,y,z)=U(x,y)\mathrm{exp}(ibz)$%
, where the stationary wave function obeys the equation
\begin{equation}
-bU=-\frac{1}{2}\nabla ^{2}U+V\left( x,y\right) U-|U|^{2}U+|U|^{4}U.
\label{NLSES}
\end{equation}

The stationary solutions are characterized by the integral power,
\begin{equation}
P=\int_{-\infty}^{+\infty}\int_{-\infty}^{+\infty}\left\vert U\left( x,y\right) \right\vert ^{2}dxdy,  \label{SP}
\end{equation}%
or its 1D counterpart. Generic results are reported below for $A=10$ in Eqs.
(\ref{VE1D}) and (\ref{VE2D}) and $b=9$ in Eq. (\ref{NLSES}).

\begin{figure}[tbp]
\begin{center}
\includegraphics[width=1\columnwidth]{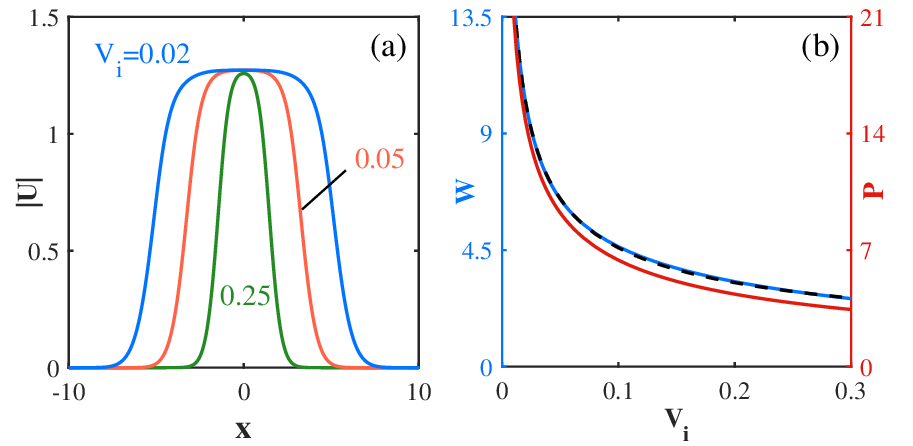}
\end{center}
\caption{(a) The profiles of 1D solitons, pinned to the potential given by Eq. (\protect\ref{VE1D}) with the single potential trough, with different values of $V_{i}$ and $x_{i}=0$. The blue, red, and green lines plot the results for $V_{i}=0.02,0.05,0.25$, respectively. (b) The half-peak width and power (the left and right vertical axes, respectively) of the 1D soliton vs. $V_{i}$. The black dashed line is the best fit of the $W\left(V_{i}\right) $ dependence to relation (\protect\ref{fit}), with $\mathrm{%
const}=1.45$. Throughout this work, we use values $A=10$ and $b=9,$ which make it possible to represent generic results.}
\label{fig1}
\end{figure}

\begin{figure}[tbp]
\begin{center}
\includegraphics[width=1\columnwidth]{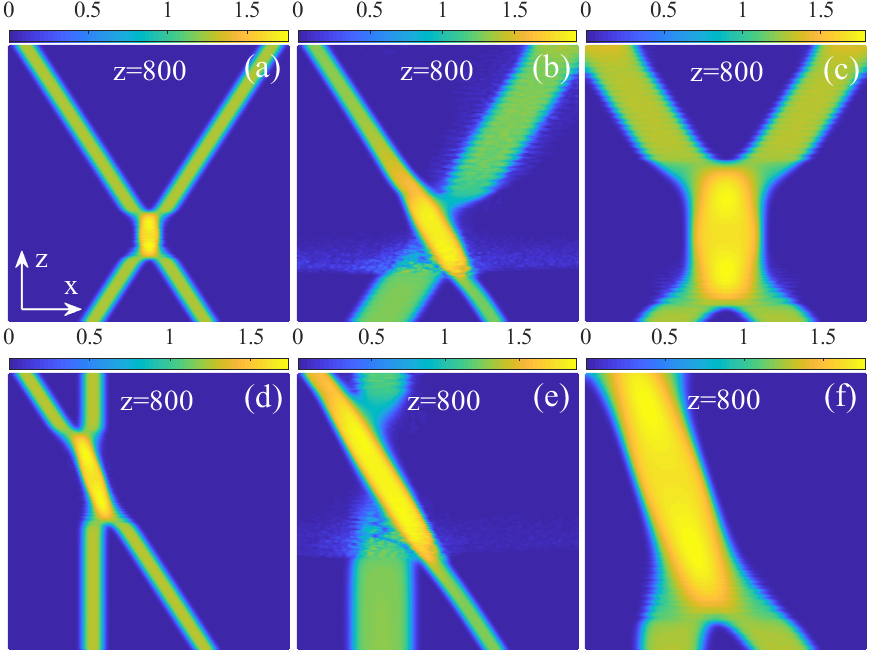}
\end{center}
\caption{Collisions of 1D solitons with $\Omega _{1}=\Omega _{2}=0.04$ in
Eq. (\protect\ref{Omega}): (a) the collision between two Gaussians solitons
steered by the troughs with $V_{1,2}=0.2$; (b) the FT-Gaussian collision,
with $V_{1}=0.2$, $V_{2}=0.02$; (c) the FT-FT collision with $V_{1,2}=0.02$.
Panels (d)--(f) are similar to (a)--(c), but for the collisions of 1D
solitons with $\Omega _{1}=0.04$, $\Omega _{2}=0$. In all panels, $\tilde{x}%
_{1}=10$, $\tilde{x}_{2}=-10$, and $|x|\leq 25$.}
\label{fig2}
\end{figure}

\begin{figure}[tbp]
\begin{center}
\includegraphics[width=1\columnwidth]{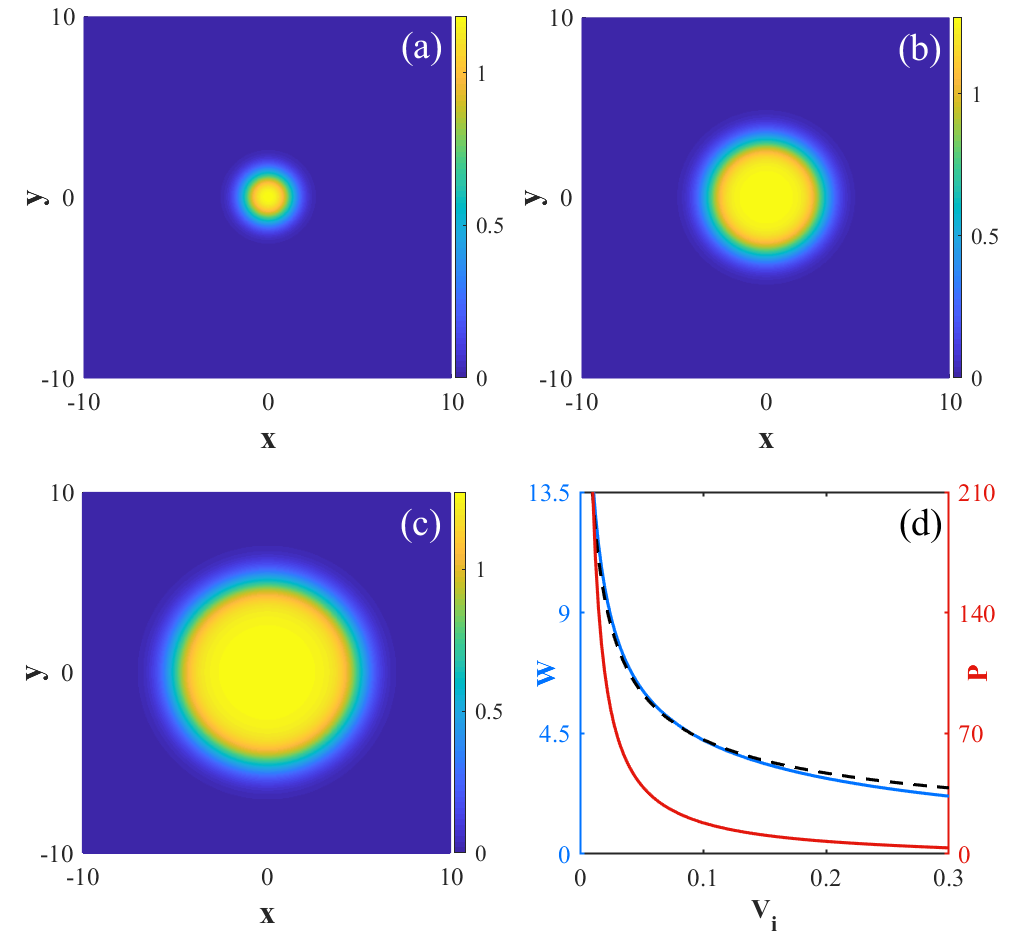}
\end{center}
\caption{(a) The contours of 2D solitons, supported by the single-trough
cylindrical potential (\protect\ref{VE2D}) with different values of $V_{i}$
and $x_{i}=0$: (a) a Gaussian soliton at $V_{i}=0.25$; (b) an FT soliton at $%
V_{i}=0.05$; (c) an FT soliton at $V_{i}=0.02$. (d) The half-peak width and
total power (the blue and red curves, respectively) of the 2D solitons vs.
the trough's inverse width $V_{i}$. The black dashed line is the best fit of
the $W\left( V_{i}\right) $ dependence to relation (\protect\ref{fit}), with
$\mathrm{const}=1.35$.}
\label{fig3}
\end{figure}

Solutions of Eq. (\ref{NLSES}) were produced by the modified
squared-operator method \cite{MSOM}. Then, their stability was tested by
means of direct simulations of the perturbed evolution in the framework of
Eq. (\ref{NLSE}), using the finite-difference algorithm.

\section{Numerical results for soliton families}

\subsection{1D solitons}

Depicted in Fig. \ref{fig1}(a) are the profiles (shown as $|U(x)|$) of 1D
solitons pinned to the potential trough with different values of inverse
width $V_{i}$. The amplitude of these solitons remains virtually constant
with the variation of $V_{i}$, which is a manifestation of the well-known
propensity of CQ solitons to maintain a constant amplitude. On the other
hand, the width of the 1D solitons naturally decreases with the increase of $%
V_{i}$, which makes the trapping potential trough narrower. For example, the
half-peak width of the soliton with $V_{i}=0.02$ and $V_{i}=0.25$ (the blue
and green lines) is $W\approx 10.3$ and $3.0$, respectively. Note that the
narrow 1D soliton with $V_{i}=0.25$ (the green line) is a Gaussian, rather
than a FT soliton. On the other hand, the 1D solitons with $V_{i}=0.02$ and $%
0.05$ (the blue and red lines, respectively) feature the FT profiles. To
further explore the relationships between the half-peak width $W$ of these
1D solitons and the inverse width $V_{i}$ of the trapping potential trough,
we present the curve of $W$ vs. $V_{i}$ in Fig. \ref{fig1}(b), in which $W$
first decreases rapidly and then slowly with the increase of $V_{i}$ [the
blue line in panel (b)]. In fact, the narrow potential trough in Eq. (\ref%
{VE1D}) determines a quasi-Gaussian shape of the trapped solitons, with the
width scaling as $W\sim V_{i}^{-1/2}$. This expectation is well
corroborated by the dashed black curves in Figs. \ref{fig1}(b) and \ref%
{fig3}(d), which display the best fit of the actual dependences $W\left(
V_{i}\right) $ to
\begin{equation}
W=\mathrm{const}\cdot V_{i}^{-1/2}.  \label{fit}
\end{equation}

Figure \ref{fig1}(b) also shows that the soliton's power $P$ (the red curve)
at first decreases rapidly and then slowly with the increase of $V_{i}$. The
similarity of the $P(V_{i})$ dependence to $W(V_{i})$ is explained by the
above-mentioned propensity of the CQ solitons to keep a constant value,
hence the total power is, roughly speaking, proportional to $W$.

The first point addressed in this work is collision between the solitons
driven by the potential troughs (\ref{VE1D}) with different slopes $\Omega
_{1}$ and $-\Omega _{2}$ in Eq. (\ref{Omega}). We start by considering the
collision between the Gaussian solitons with $\Omega _{1,2}=0.04$, which is
displayed in Fig. \ref{fig2}(a), where the Gaussians collide elastically,
continuing their undisturbed motion after the collision. The random noise with strength of $1\%$ of the amplitude is added in all the numerical simulations.

\begin{figure}[tbp]
\begin{center}
\includegraphics[width=1\columnwidth]{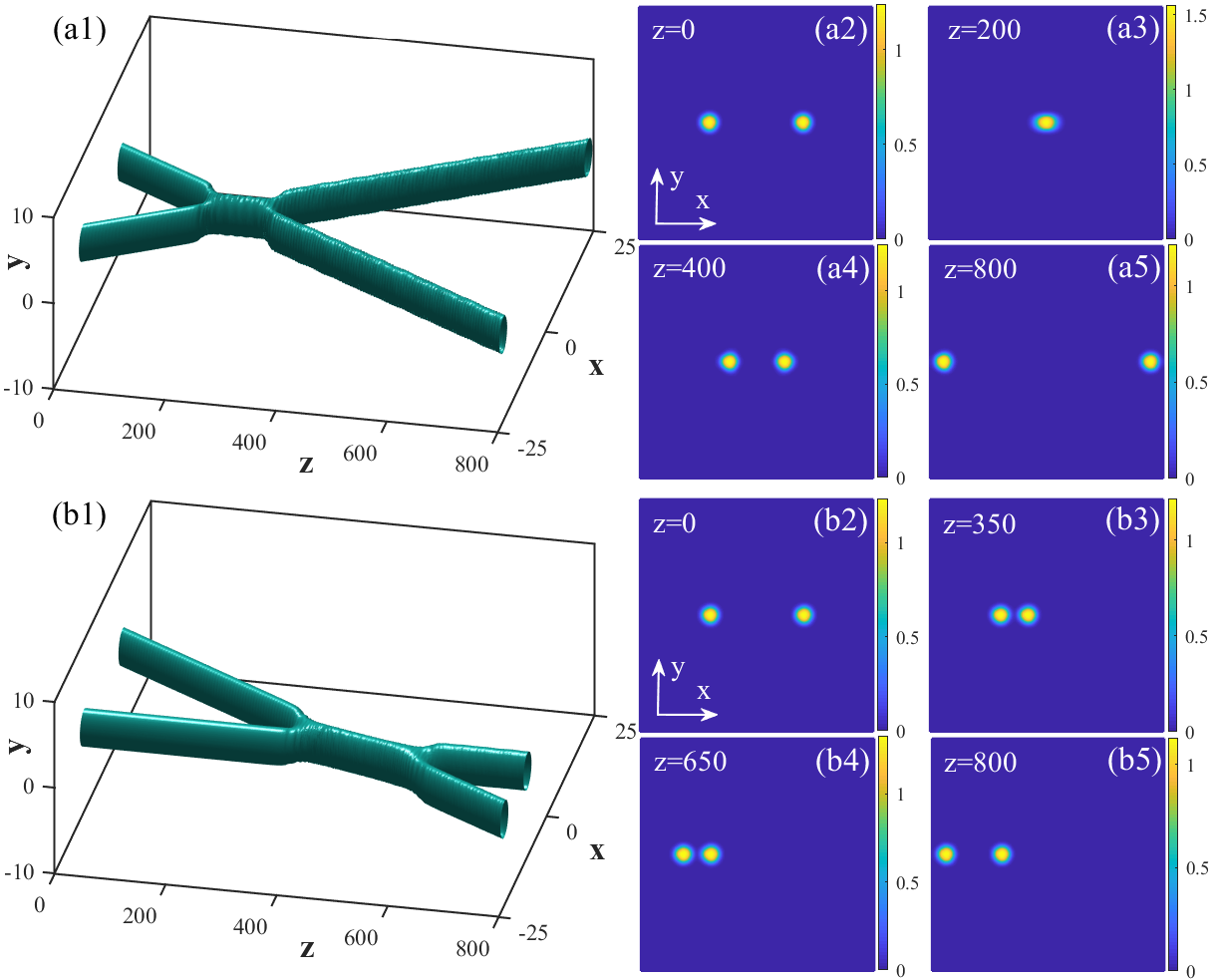}
\end{center}
\caption{(a1) Collisions of 2D Gaussian solitons with ($V_{1}=0.2, V_{2}=0.2$) and ($\Omega_{1}=0.04, \Omega_{2}=0.04$). Contours of the collision in panel (a1) at different values of the propagation distance $z$: (a2) at $z=0$, (a3) at $z=200$, (a4) $z=400$, (a5) $z=800$. Panels (b1)--(b5) display the results similar to those in panels (a1)--(a5), but for ($\Omega_{1}=0.04,\Omega_{2}=0$). In all panels, $\tilde{x}_{1}=10$, $\tilde{x}_{2}=-10$, $\left\vert x,y\right\vert \leq 25.$}
\label{fig4}
\end{figure}

The collision between an FT soliton and a Gaussian with a large amplitude,
that are steered by the troughs with the same slopes, corresponding to $%
\Omega _{1,2}=0.04$ in Eq. (\ref{Omega}), is presented in Fig. \ref{fig2}%
(b), which demonstrates that the solitons again pass through each
quasi-elastically, generating some weak radiation field. Further, the
collision of 1D FT solitons with $\Omega _{1,2}=0.04$ is displayed in Fig. %
\ref{fig2}(c) where they also continue essentially undisturbed motion after
the collision.

Another set of collisions between the same pairs of the 1D solitons as in
Figs. \ref{fig2}(a-c), but steered by the troughs with slopes $\Omega
_{1}=0.04$ and $\Omega _{2}=0$ in Eq. (\ref{Omega}), is plotted in Figs. \ref%
{fig2}(d-f). Due to the Galilean invariance of the underlying nonlinear Schr%
\"{o}dinger equation in the free space (without the trough potentials),
these results are essentially the same as for $\Omega _{1}=\Omega _{2}=0.02$%
, see Eq. (\ref{Omega}).

It is seen that the quasi-elastic Gaussian-Gaussian and FT-Gaussian
collisions, which are displayed in panels (d,e), are quite similar to those
presented in (a,b), respectively. On the other hand, the \textit{inelastic}
FT-FT collision, revealed by Fig. \ref{fig2}(f), is completely different
from its counterpart in panel (c), as here the colliding FT solitons merge
into a single one. Note that the merger implies strong \textit{spontaneous
symmetry breaking} (SSB) in the system \cite{SSB}, as the single emerging
soliton spontaneously chooses one potential trough out of two. Below, it is
demonstrated that the merger of colliding 2D FT solitons also implies SSB,
see Figs. \ref{fig6}(b1-b5). The possibility of the
latter outcome of the FT-FT collision is explained by the fact that a
Gaussian soliton passing through another Gaussian soliton or a FT soliton keeps its individual
identity, thus re-emerging in the original form after the collision. On the
other hand, when colliding FT solitons stay overlapped in the course of long
copropagation, the wave field may undergo homogenization, preventing the
subsequent separation of the overlapped state into the original FT-soliton
pair. The same argument explains the possibility of the merger of colliding
2D FT solitons, see Figs. \ref{fig6}(b1-b6) below.

\subsection{2D solitons and a summary of the results}

The contours of 2D solitons with different values of $V_{i}$ are displayed
in Figs. \ref{fig3}(a)--(c), where their amplitude are the same, in
agreement with the above-mentioned property of the CQ model, while the width
naturally increases with the decrease of the inverse width $V_{i}$ of the 2D
trapping potential, similar to the 1D results. The curves of the half-peak
width $W$ and soliton power $P$ vs. $V_{i}$ for the 2D soliton families are
plotted by the left blue and red lines, respectively, in Fig. \ref{fig3}(d).
Similar to the 1D case, both $W$ and $P$ decrease first rapidly and then
slowly with the increase of $V_{i}$.

\begin{figure}[tbp]
\begin{center}
\includegraphics[width=1\columnwidth]{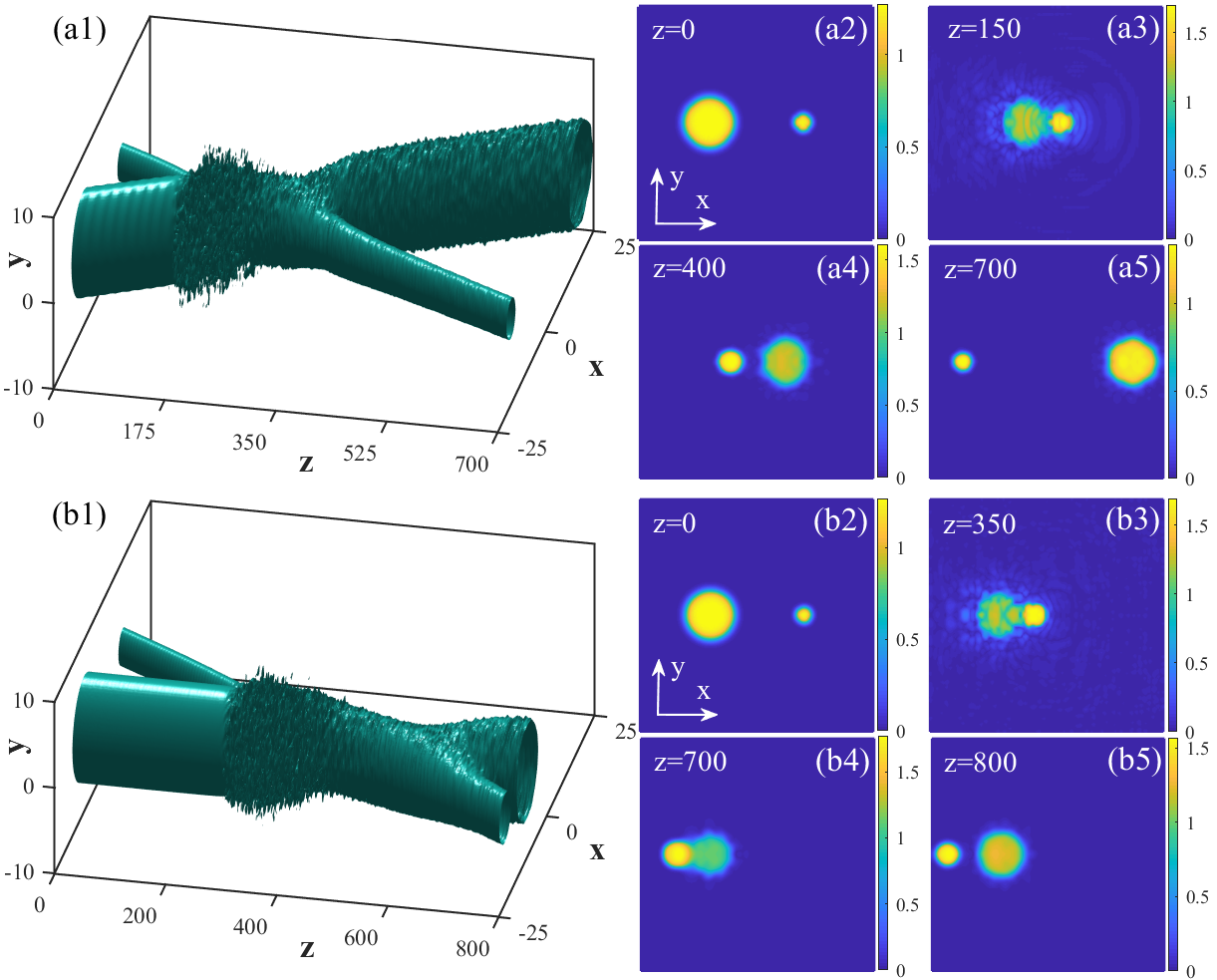}
\end{center}
\caption{(a1) The 2D collision of a flat-top soliton and a Gaussian soliton with $%
V_{1}=0.2$, $V_{2}=0.02$ at $\Omega _{1}=0.04$, $\Omega _{2}=0.04$. The
contours of the collision in panel (a1) at different values of the
propagation distance $z$: (a2) $z=0$; (a3) $z=150$; (a4) $z=400$; (a5) $%
z=700 $. The results displayed in panels (b1)--(b5) are similar to those in
panels (a1)--(a5), but for $\Omega _{1}=0.04$, $\Omega _{2}=0$. In all
panels, $\tilde{x}_{1}=10$, $\tilde{x}_{2}=-10$, $x,y\in [-25,+25]$.}
\label{fig5}
\end{figure}

Next we address collisions of the 2D solitons. The collisions of two 2D
Gaussian solitons with $\left( \Omega _{1}=0.04,\Omega _{2}=0.04\right) $ and $%
\left( \Omega _{1}=0.04,\Omega _{2}=0\right) $ in Eq. (\ref{Omega}) are
displayed in Figs. \ref{fig4}(a1)--(a5) and \ref{fig4}(b1)--(b5),
respectively. Similar to the 1D results presented above in Figs. \ref{fig2}%
(a,d), Fig. \ref{fig4}(a1) shows that the collision of two 2D Gaussian solitons are
quasi-elastic, including a temporary formation of a fused \textquotedblleft
neck".

Collisions of FT solitons with Gaussian solitons, for the steering parameters $%
\left( \Omega _{1}=\Omega _{2}=0.04\right) $ and $\left( \Omega
_{1}=0.04,\Omega _{2}=0\right) $ (alias $\Omega _{1}=\Omega _{2}=0.02$, as
mentioned above) in Eq. (\ref{Omega}), are displayed in Figs. \ref{fig5}%
(a1)--(a5) and \ref{fig5}(b1)--(b5), respectively, where the amplitude of
the Gaussian soliton is larger than the FT's amplitude. Similar to the 1D results
shown above in Figs. \ref{fig2}(b,e), Figs. \ref{fig5}(a1) exhibit that the
collisions remain overall quasi-elastic, accompanied by the generation of
radiation.

\begin{figure}[tbp]
\begin{center}
\includegraphics[width=1\columnwidth]{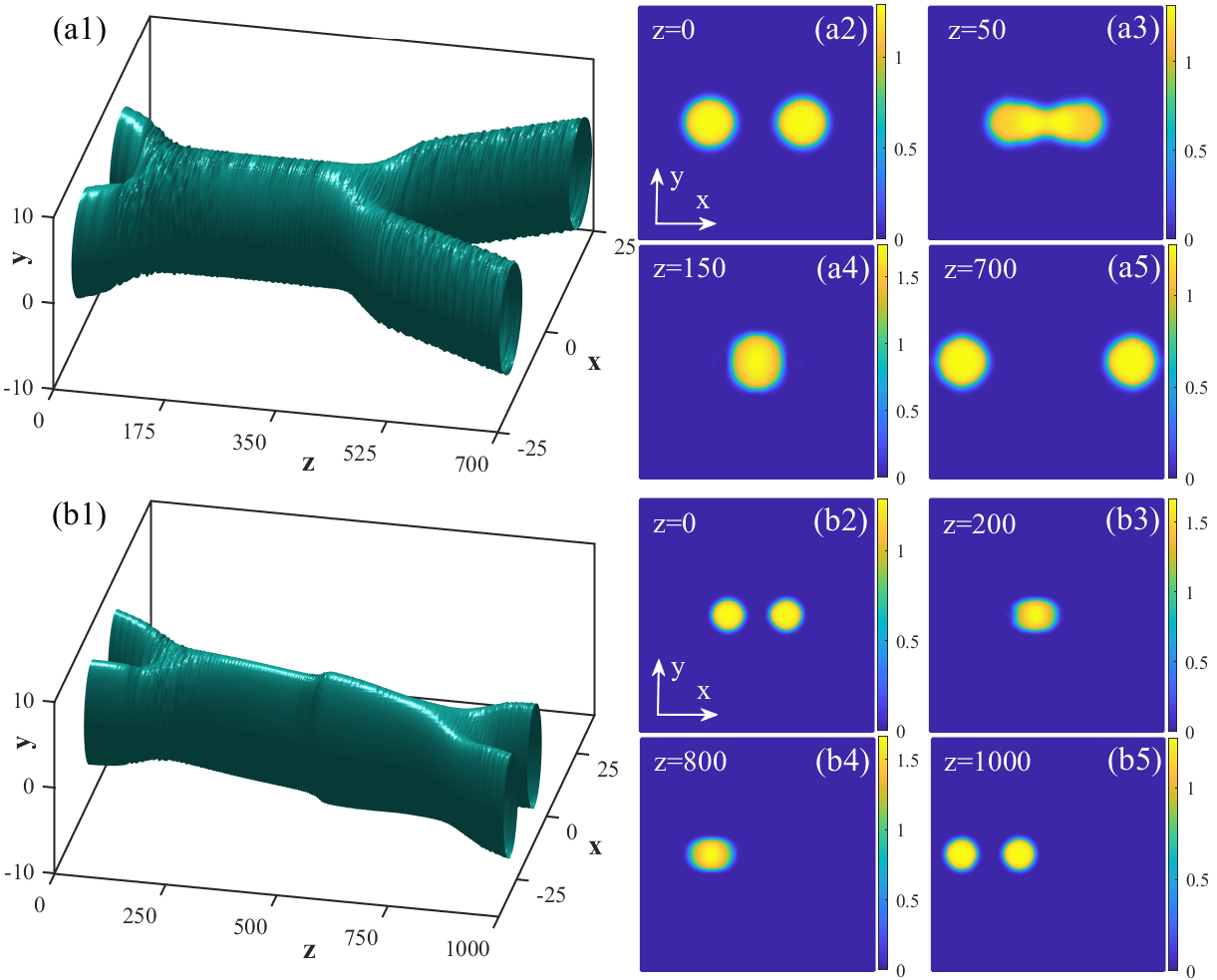}
\end{center}
\caption{{(a1) The 2D collision of two flat-top solitons with $V_{1}=0.02$, $%
V_{2}=0.02$ at $\Omega _{1}=0.04$, $\Omega _{2}=0.04$. The contours of the
collision in panel (a1) at different values of the propagation distance $z$:
(a2) $z=0$; (a3) $z=200$; (a4) $z=400$; (a5) $z=800$. The results displayed
in panels (b1)--(b5) are similar to those in panels (a1)--(a5) but for $%
\Omega _{1}=0.04$, $\Omega _{2}=0$. $\tilde{x}_{1}=10$, $\tilde{x}_{2}=-10$,
$x,y\in [-25,+25]$ for panels (a2)--(a5); $x,y\in [-40,+40]$ for panels (b2)--(b5).}}
\label{fig6}
\end{figure}

The collisions of 2D FT solitons with $\left( \Omega _{1}=0.04,\Omega
_{2}=0.04\right) $ and $\left( \Omega _{1}=0.04,\Omega _{2}=0\right) $ are
presented in Figs. \ref{fig6}(a1)--(a5) and \ref{fig6}(b1)--(b5),
respectively. Similar to the results reported in Figs. \ref{fig2}(c) and
(f), Figs. \ref{fig6}(a1) and (b1) exhibit, severally, examples of the
elastic and inelastic collisions (alias mutual passage and merger of the
colliding solitons).

\begin{figure}[tbp]
\begin{center}
\includegraphics[width=1\columnwidth]{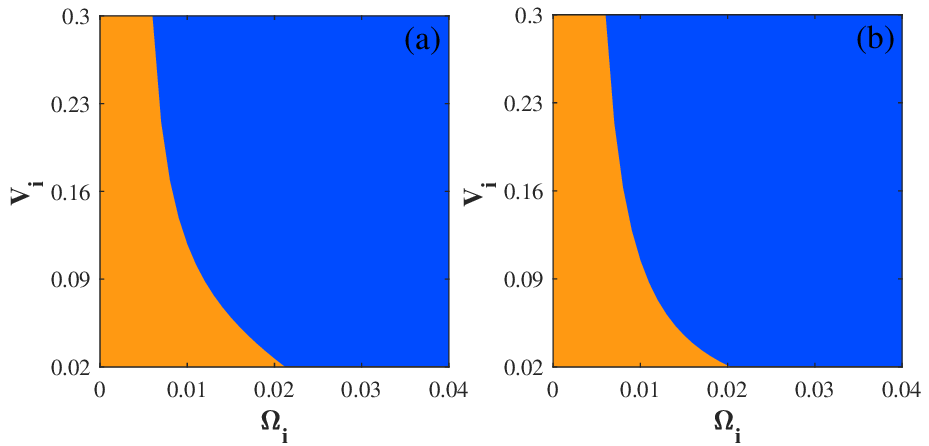}
\end{center}
\caption{Outcomes of collisions between two 1D (a) or 2D (b) solitons, defined as per Eqs. (\protect\ref{Omega}) with $\Omega _{1}=\Omega_{2}\equiv \Omega _{{\protect\normalsize i}}$ and equal parameters $V_{i}$ in Eqs. (\protect\ref{VE1D}) or (\protect\ref{VE2D}): merger or quasi-elastic passage, in yellow and blue areas, respectively. The parameter charts are not extended to $V_{i}<0.02$, as for so broad troughs the numerical solution cannot produce two far-separated FT solitons.}
\label{fig7}
\end{figure}

The results produced by the systematic simulations of the soliton-soliton
collisions in the 1D and 2D systems are summarized in the parameter charts
plotted in Figs. \ref{fig7}(a) and (b), respectively. The charts display the
boundaries between the quasi-elastic passage and merger outcomes of the
collisions. Naturally, the increase of the tilt difference, $2\Omega _{i}$,
leads to the transition from the merger to passage. In agreement with the
above-mentioned argument, the merger occurs, chiefly, at parameter values at
which the colliding solitons are ones of the FT type, similar to what is
demonstrated above in Figs. \ref{fig2}(f) and \ref{fig6}(b1-b5).
Nevertheless, collisions between slowly moving Gaussian solitons also lead to the
merger.

Accordingly, the increase of $V_{i}$, making the solitons narrower, also
leads to a gradual transition from the merger to passage.

\section{Conclusion}

\label{sec4}

In this work, we have demonstrated that\ 1D and 2D Gaussian-shaped and FT
(flat-top) solitons, steered by the intersecting potential troughs added to
the nonlinear Schr\"{o}dinger equation with the CQ (cubic-quintic)
nonlinearity, demonstrate quasi-elastic or inelastic collisions (mutual
passage or merger into a single soliton, respectively). In the latter case,
the soliton produced by the merger is trapped in the single steering trough,
which implies strong spontaneous symmetry breaking. Namely,
Gaussian-Gaussian and Gaussian-FT collisions turn out to be quasi-elastic,
while the FT-FT collisions may be either quasi-elastic or lead to the
symmetry-breaking merger into a single FT. In addition to this, it is found
that the Gaussian-FT collisions, remaining overall quasi-elastic, give rise
to the generation of weak radiation fields.

The setting elaborated in this work may be used for the design of
all-optical data-processing schemes, with individual spatial solitons
carrying unit bits. Another extension of the work is possible for the effectively 1D
Gross-Pitaevskii equation (GPE) for atomic Bose-Einstein condensates, which
features the combination of cubic self-repulsion and quadratic
self-attraction. The quadratic term accounts for the Lee-Huang-Yang
correction to the mean-field approximation, which gives rise to \textit{%
quantum droplets} \cite{PA,QD1,QD2}. The GPE with the quadratic-cubic nonlinearity
also gives rise to families of Gaussian-shaped and FT solitons. Collisions
between such solitons in the free space may be elastic or inelastic,
depending on parameters \cite{Astra}. Further, an interesting extension of this work maybe the effect of the phase difference between
colliding solitons on the outcome of the collision. In particular, it is expected that the phase difference of $\pi$ will reverse the
sign of the interaction force.

\section*{ACKNOWLEDGEMENT}
This work was supported by National Natural Science Foundation of China (62205224, 61827185, 11774068), Guangdong Basic and Applied Basic Research Foundation (2023A1515010865), Guangzhou Science and Technology Plan Project (2025A04J4068), and Israel Science Foundation (1695/22).

\section*{DECLARATION OF COMPETING INTEREST}
The authors have no conflicts to disclose.

\section*{AUTHOR CONTRIBUTIONS}
\textbf{Liangwei Zeng:} Conceptualization (equal); Investigation (equal); Methodology (equal); Writing - original draft (lead); Writing - review \& editing (equal). \textbf{Boris A. Malomed:} Formal analysis (equal); Methodology (equal); Writing - review \& editing (equal). \textbf{Dumitru Mihalache:} Formal analysis (equal); Methodology (equal); Writing - review \& editing (equal). \textbf{Jingzhen Li:} Methodology (equal); Writing - review \& editing (equal). \textbf{Xing Zhu:} Conceptualization (equal); Investigation (equal); Methodology (equal); Writing - review \& editing (equal).

\section*{DATA AVAILABILITY STATEMENT}
The data that support the findings of this study are available
from the corresponding author upon reasonable request.

\end{document}